# Exploring linearity of deep neural network trained QSM: QSMnet+

Woojin Jung[1], Jaeyeon Yoon[1], Sooyeon Ji[1], Joon Yul Choi[2], Jae Myung Kim[3], Yoonho Nam[4], Eung Yeop Kim[5], and Jongho Lee[1,*]

**Author Affiliation:**

[1]Laboratory for Imaging Science and Technology, Department of Electrical and Computer Engineering, Seoul National University, Seoul, Korea

[2]Epilepsy Center, Neurological Institute, Cleveland Clinic, OH, USA

[3]Department of Electrical and Computer Engineering, Seoul National University, Seoul, Korea

[4]Department of Radiology, Seoul St. Mary's Hospital, College of Medicine, The Catholic University of Korea, Seoul, Korea

[5]Department of Radiology, Gil Medical Center, Gachon University College of Medicine, Incheon, Korea

**Corresponding Author:**

Jongho Lee[1]

Department of Electrical and Computer Engineering, Seoul National University

Building 301, Room 1008, 1 Gwanak-ro, Gwanak-gu, Seoul, Korea

Tel: 82-2-880-7310

E-mail: jonghoyi@snu.ac.kr

**Abstract**

Recently, deep neural network-powered quantitative susceptibility mapping (QSM), QSMnet, successfully performed ill-conditioned dipole inversion in QSM and generated high-quality susceptibility maps. In this paper, the network, which was trained by healthy volunteer data, is evaluated for hemorrhagic lesions that have substantially higher susceptibility than healthy tissues in order to test "linearity" of QSMnet for susceptibility. The results show that QSMnet underestimates susceptibility in hemorrhagic lesions, revealing degraded linearity of the network for the untrained susceptibility range. To overcome this limitation, a data augmentation method is proposed to generalize the network for a wider range of susceptibility. The newly trained network, which is referred to as QSMnet$^+$, is assessed in computer-simulated lesions with an extended susceptibility range (-1.4 ppm to +1.4 ppm) and also in twelve hemorrhagic patients. The simulation results demonstrate improved linearity of QSMnet$^+$ over QSMnet (root mean square error of QSMnet$^+$: 0.04 ppm *vs.* QSMnet: 0.36 ppm). When applied to patient data, QSMnet$^+$ maps show less noticeable artifacts to those of conventional QSM maps. Moreover, the susceptibility values of QSMnet$^+$ in hemorrhagic lesions are better matched to those of the conventional QSM method than those of QSMnet when analyzed using linear regression (QSMnet$^+$: slope = 1.05, intercept = -0.03, $R^2$ = 0.93; QSMnet: slope = 0.68, intercept = 0.06, $R^2$ = 0.86), consolidating improved linearity in QSMnet$^+$. This study demonstrates the importance of the trained data range in deep neural network-powered parametric mapping and suggests the data augmentation approach for generalization of network. The new network can be applicable for a wide range of susceptibility quantification.

**Introduction**

Deep learning has been widely applied for MRI reconstruction and processing and has demonstrated promising outcomes (Akcakaya et al., 2019; Bollmann et al., 2019; Cohen et al., 2018; Hammernik et al., 2018; Han et al., 2018; Knoll et al., 2019; Quan et al., 2018; Ronneberger et al., 2015; Yoon et al., 2018). Recently, it has been observed that the performance of a deep neural network is highly dependent on the training dataset. For example, if a network is trained with a specific range of training dataset (e.g. image SNR of 80), the performance of the network can be substantially degraded for an input outside of this range (e.g. a test image SNR of 20) (Høy et al., 2019; Knoll et al., 2019). These types of input data are referred to as "out-of-distribution data" for a given network and have of great importance in network generalization (Amodei et al., 2016; Goodfellow et al., 2015).

One particular field of MRI that requires a special consideration for out-of-distribution data is parametric mapping, which quantitatively measures such MR parameters as T1, T2, diffusion, susceptibility, etc (Bertleff et al., 2017; Cohen et al., 2018; Lee et al., 2018; Yoon et al., 2018). When a network is trained for quantitative parametric mapping, it is often trained with a range of the parameter due to the limited availability of training data. When the network is applied for an input dataset, however, it may fall outside of the training data range, potentially generating unwanted outcomes. Hence, it is necessary to develop an approach to compensate for this issue.

Quantitative susceptibility mapping (QSM) was introduced as a quantitative approach for measuring magnetic susceptibility using MRI (de Rochefort et al., 2008; Liu et al., 2009; Shmueli et al., 2009). Because of the sensitivity to susceptibility sources such as hemoglobin, myelin and ferritin, QSM has been applied for several brain disorders including brain hemorrhage, multiple sclerosis, and Parkinson's disease (Chen et al., 2014; Kim et al., 2018; Liu et al., 2015; Sung et al., 2019; Wang and Liu, 2015). However, the key process of QSM reconstruction, which is referred to as dipole inversion, is an ill-conditioned problem and has been an active area of research. Several methods that utilize regularization have been proposed to improve the quality of QSM maps (Bilgic et al., 2014; de Rochefort et al., 2010; Li et al., 2015; Li et al., 2011; Liu et al., 2011; Liu et al., 2009; Schweser et al., 2012; Wharton et al., 2010). Despite these improvements, reconstruction errors such as streaking artifacts were still pertained. As an alternative approach, a method, COSMOS, that utilized multiple head orientation data was proposed and generated a high-quality susceptibility map at the cost of a long scan time (five times or longer when compared to a single orientation scan) and patient discomfort (Liu et al., 2009).

Recently, a deep neural network has been proposed as a tool for the QSM reconstruction (Bollmann et al., 2019; Chen et al., 2019b; Liu and Koch, 2019b; Wei et al., 2019; Yoon et al., 2018). When a deep neural network was trained using local field maps and COSMOS QSM maps, the network generated a COSMOS-quality QSM map from a single head orientation data (Yoon et al., 2018). This network, which is referred to as QSMnet, has been applied to a few patient data and demonstrated successful delineation of lesions despite the fact that the network was trained only by healthy volunteers (Yoon et al., 2018). However, no quantitative analysis was performed for the patient results. Since the magnetic susceptibility range of healthy volunteers can be different from that of patients, the issue of out-of-distribution data may exist in the susceptibility mapping. Therefore, exploring the network-generated susceptibility estimations for out-of-distribution data is of importance. Finding this characteristic can be considered as a test of the "linearity" in QSM since we are exploring a linear relationship between true susceptibility and network-generated susceptibility.

In this study, we investigated the linearity of QSMnet for out-of-distribution data. Furthermore, we suggested a data augmentation method to enhance the linearity. The proposed augmentation method was validated using simulation and experimental data. The newly trained network, which is referred to as QSMnet$^{+}$, is available at https://github.com/SNU-LIST/QSMnet.

## Methods

**[Susceptibility distribution in healthy volunteers]**

As the first step, the magnetic susceptibility distribution of healthy volunteers was explored using the experimental data in the training dataset of QSMnet. The dataset had five healthy volunteer scans. Each scan had five head orientation data, which were used for COSMOS QSM reconstruction. Non-brain regions were excluded using a brain mask (BET, FSL, Oxford, UK; Smith, 2002). Then, the susceptibility values of the five brains were counted to generate a histogram of the susceptibility distribution.

**[Data augmentation for improved linearity]**

To improve linearity in the susceptibility range, a data augmentation method was developed by expanding the susceptibility range of the training dataset (Fig. 1). The extended susceptibility range was generated as follows:

$$\tilde{\chi} = \Lambda \cdot \chi \qquad \text{[Eq. 1]}$$

$$\tilde{\delta} = FDF^{-1}\tilde{\chi} \qquad \text{[Eq. 2]}$$

where $\tilde{\chi}$ is a susceptibility map (in ppm) generated by the data augmentation, $\chi$ is an input susceptibility map (in ppm), $\Lambda$ is a scaling map defined below, $\cdot$ is a symbol for voxel-wise multiplication, $\tilde{\delta}$ is a field map generated by the data augmentation, $F$ is a Fourier transform matrix, and $D$ is a dipole convolution matrix. The scaling map, $\Lambda$, was defined as written below:

$$\Lambda(x) = \begin{cases} \lambda & if \ x \in P \\ 1 & if \ x \notin P, \end{cases} \quad \text{where } \lambda \geq 1 \qquad \text{[Eq. 3]}$$

where $x$ is a voxel, $\lambda$ is a scaling factor, and P represents a set of the voxels where $\lambda$ is applied. In our method, P was defined for voxels with positive susceptibility values in the input susceptibility map, scaling only positive susceptibility sources (Fig. 1b; see Supplementary Fig. S1-3 and Table. S1 for other conditions of P tested). The scaling factor ($\lambda$) was defined to be greater than or equal to 1 so that the augmentation increased the range of the original susceptibility map. To generate a symmetric susceptibility distribution (Fig. 1e), additional datasets were generated by inverting the sign of the input and scaled susceptibility maps (Figs. 1c and d).

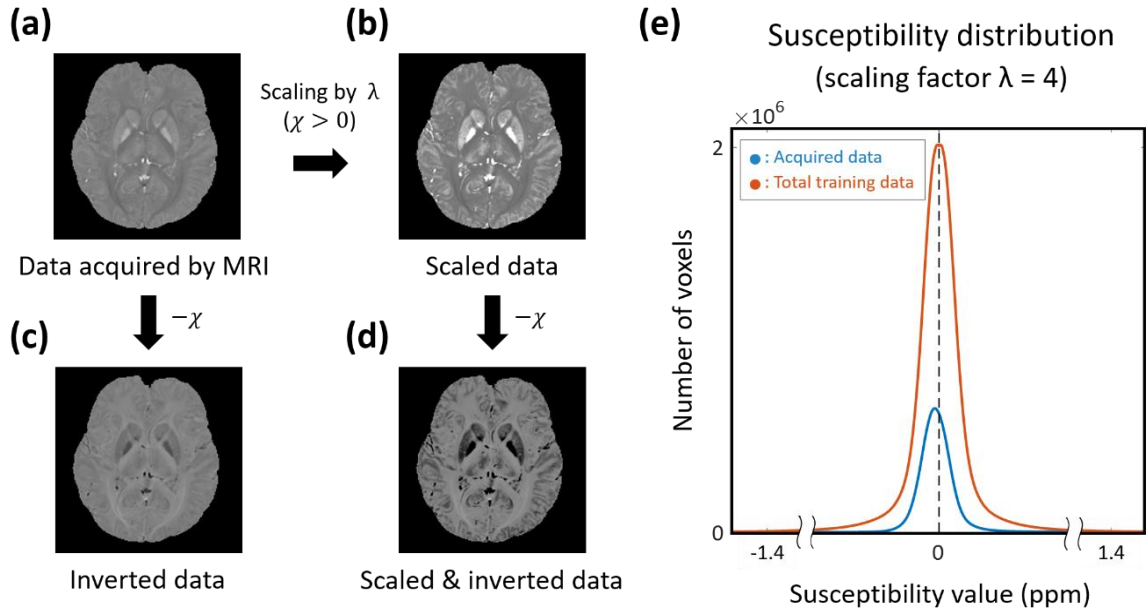

*Figure 1.* Summary of the data augmentation method: (a) Data acquired by MRI, (b) scaled data where positive susceptibility values ($\chi > 0$) are increased by a scaling factor $\lambda$ using the augmentation method, (c) sign-inverted data, and (d) scaled and sign-inverted data. (e) The plot on the right shows the susceptibility distributions of the acquired data (blue) and total training data (sum of the four data; red). The distribution of the acquired data (blue) shows asymmetry whereas that of the total training data shows symmetry over a wider susceptibility range.

This data augmentation process was applied to the COSMOS reconstructed susceptibility maps of the five subjects in the QSMnet training dataset. For each subject, four additional susceptibility maps of different head orientations were generated by applying a rotation matrix to the COSMOS reconstructed susceptibility map in order to increase the diversity of head orientation in the training dataset. The rotation angle was randomly chosen between -30° to 30° relative to $B_0$. Then, the proposed data augmentation process was repeated to all orientation data, generating a total of 25 "scaled" susceptibility maps and corresponding local field maps. The same process was also repeated for the sign inverted susceptibility maps.

In summary, we had four different types of training datasets: the original dataset from QSMnet, the scaled dataset, the inverted dataset, and the scaled and inverted dataset. The total training dataset had 100 susceptibility maps and corresponding local field maps. This dataset was used to generate a new neural network, which is referred to as QSMnet[+], that improves the linearity of QSMnet in the susceptibility range.

**[Deep neural network]**

For a deep neural network, the QSMnet structure, which is a 3D U-net architecture, was utilized with a modification of ReLU to leaky ReLU (Figure S2; Yoon et al., 2018). The encoder part of the network consisted of 5 blocks. Each block contained two 5 x 5 x 5 convolution layers. Each layer was followed by batch normalization and leaky ReLU (slope = 0.1) (Maas et al., 2013). Then, a 2 x 2 x 2 max-pooling layer with strides of two was performed except for the last block. The number of channels was 32 in the first layer and it doubled for each subsequent layer. The decoder consisted of 4 blocks. Each block contained a 2 x 2 x 2 deconvolution layer, followed by two 5 x 5 x 5 convolution layers. Batch normalization and leaky ReLU (slope = 0.1) were performed after each convolutional layer. with compressing the number of channels by half. In order to forward the feature information from the encoder to the decoder, four feature concatenations were applied between corresponding encoder block and decoder block. In the last layer, a 1 x 1 x 1 convolution was performed with reducing the number of output channels to 1.

The loss function in QSMnet, which had model loss ($loss_{Model}$), L1 loss ($loss_{L1}$), and gradient difference loss ($loss_{Gradient}$) (Yoon et al., 2018) was modified by defining the model loss as follows:

$$loss_{Model} = ||M \cdot (\delta - d * \chi)||_1 \qquad [Eq. 7]$$

where M is a brain mask, $\delta$ is an input local field, d is a dipole kernel, and $\chi$ is an output susceptibility map. The other two losses were unchanged:

$$loss_{L1} = ||(\chi - y)||_1 \qquad [Eq. 8]$$

$$loss_{Gradient} = \sum_{i=x,y,z} ||\nabla\chi|_i - |\nabla y|_i| \qquad [Eq. 9]$$

where y is a label. The total loss was defined as the weighted sum of the three losses with empirically determined weighting parameters of 0.5, 1, and 0.1 for the model loss, L1 loss, and gradient different loss, respectively.

The learning rate was exponentially decayed from $10^{-3}$ at every 600 steps with decay factor of 0.95. The training process was stopped at 25 epochs. All the other settings and hyper-parameters were the same as QSMnet: Xavier initializer (Glorot et al., 2011), RMSProp for optimization, and batch size of 12. Tensorflow (Rampasek and Goldenberg, 2016) and an Nvidia GTX 1080Ti GPU (Nvidia Crop., Santa Clara, CA) were utilized for network training and inference.

**[Network training]**

The training dataset was normalized for efficient convergence of the gradient descent method (LeCun et al., 1998). For normalization, the mean and standard deviation were calculated in the training dataset of the 100 susceptibility maps. Then, the susceptibility maps were normalized to have a mean of 0 and a standard deviation of 1 (LeCun et al., 1998). The local field maps were also normalized using the same process. When generating an output susceptibility map, the outcome of the neural net was re-scaled using the normalization parameters to restore the susceptibility unit to ppm.

When training the network, the normalized input (local field map) and the normalized output (susceptibility map) were divided into 3D patches with a size of 64 x 64 x 64 voxels. The patch was generated with a 66% overlap. Overall, a total of 33,600 patches was used for training.

To investigate the effect of $\lambda$ in QSMnet$^+$, four different augmentation datasets were created using $\lambda$ of 1, 2, 3, and 4. The maximum $\lambda$ value of 4 was chosen such that the susceptibility distribution of the training dataset sufficiently covered the susceptibility value of the fully deoxygenated blood, which was approximately 1.4 ppm (see Discussion) (Chang et al., 2016; Haacke et al., 1997; Zborowski et al., 2003). The four different QSMnet$^+$ networks were labeled as $\text{QSMnet}^+_{\lambda=1}$, $\text{QSMnet}^+_{\lambda=2}$, $\text{QSMnet}^+_{\lambda=3}$, and $\text{QSMnet}^+_{\lambda=4}$.

For comparison, QSMnet was re-trained using the new activation function, loss function, learning rate and normalization process to avoid complications from the modifications in this study. Hence, differences between the results of QSMnet and QSMnet$^+$ were primarily from the difference in the data augmentation.

**[Datasets for network evaluation]**

The evaluation of QSMnet and QSMnet$^+$ was performed using datasets of five healthy volunteers and twelve patients with hemorrhage. The five healthy volunteer test sets were from Yoon et al., (Yoon et al., 2018). Each volunteer dataset consisted of five local field maps with different head orientations and a corresponding COSMOS QSM map. The dataset was scanned at 3T (Skyra, SIEMENS, Erlangen, Germany) with the following scan parameters: 3D single-echo GRE data were acquired with FOV = 176 x 176 x 100 mm$^3$, voxel size = 1 x 1 x 1 mm$^3$, TR = 33 ms, TE = 25 ms, bandwidth = 100 Hz/pixel, and flip angle = 15°.

The twelve patient test sets were acquired to investigate the effects of the out-of-distribution data (i.e. high susceptibility values) in hemorrhage (3T, 12 patients using Skyra, SIEMENS, Erlangen, Germany; 2 patients using Ingenia, PHILIPS, Best, Netherlands). All scans were approved by IRB. The data scanned in SIEMENS were acquired with the same scan parameter as the healthy volunteer data except for FOV of 192 x 192 x 80 mm$^3$. The sequence parameters for PHILIPS were as follows: FOV = 220 x 220 x 144 mm$^3$, voxel size = 0.5 x 0.5 x 2 mm$^3$, TR = 36 ms, TE = 25 ms, bandwidth = 255 Hz/pixel, and flip angle = 17°. The resolution of the PHILIPS data was interpolated to 1 x 1 x 1 mm$^3$ by zero-padding in slice and truncating in in-plane in the Fourier domain. For preprocessing, a brain mask was extracted from the magnitude image using BET (FSL, Oxford, UK; Smith, 2002). Within the brain mask, the phase image was unwrapped by Laplacian phase unwrapping (Li et al., 2011). Finally, a local field map was generated by removing a background field using V-SHARP (Wu et al., 2012).

[Simulated lesion]

To investigate the linearity of QSMnet and QSMnet$^+$ for a susceptibility range, a simulated local field map was generated by combining the local field maps of a healthy volunteer and a simulated lesion. The healthy volunteer data was chosen from the test sets. The field map of the simulated lesion was generated as follows: First, a hemorrhagic lesion was manually segmented in the magnitude image of one of the patients. A susceptibility value was assigned to the lesion. Then, Gaussian noise with a mean of zero and a standard deviation of the half of the assigned susceptibility value was added to the lesion. This simulated lesion was generated 17 times for lesion susceptibility values from -1.4 ppm to 1.4 ppm in the step size of

0.2 ppm. The maximum susceptibility of 1.4 ppm represented the susceptibility of fully deoxygenated blood. From the lesion susceptibility map, a local field map was generated by dipole convolution. After that, this local field map was added to that of the healthy volunteer.

From the final local field map, susceptibility maps were reconstructed using STAR-QSM (Wei et al., 2015), QSMnet, $QSMnet^+_{\lambda=1}$, $QSMnet^+_{\lambda=2}$, $QSMnet^+_{\lambda=3}$, and $QSMnet^+_{\lambda=4}$. Note that $QSMnet^+_{\lambda=4}$ was considered as a default network and is referred to as $QSMnet^+$ if not noted. Reconstruction results were evaluated by averaging the susceptibility values in the lesion. Root mean square error (RMSE) between the assigned and reconstructed susceptibility values in the lesion ROI was calculated.

[Patients]

The twelve hemorrhage patients were reconstructed using two model-based QSM (MEDI (Liu et al., 2011) and STAR-QSM (Wei et al., 2015)), QSMnet and $QSMnet^+$. The data from two patients, which showed the diamagnetic susceptibility in the lesions were excluded in analysis. The resulting maps were visually inspected for reconstruction quality. For a region of interest (ROI) analysis, the largest hemorrhagic lesion in each patient was manually segmented in the GRE magnitude images. Linear regression was applied between STAR-QSM *vs.* QSMnet and STAR-QSM *vs.* $QSMnet^+$ for the mean susceptibility values of the lesion ROIs. Finally, $R^2$ values were calculated to demonstrate the goodness of fit between the measurements.

[Health controls]

To demonstrate that the proposed augmentation approach sustained the quality of QSM maps in healthy volunteers, the five healthy volunteer test sets were processed using the new networks. The quantitative metrics, peak signal-to-noise-ratio (pSNR), normalized root mean squared error (NRMSE), high-frequency error norm (HFEN), and structure similarity index (SSIM), were calculated with the COSMOS QSM map as a reference (Langkammer et al., 2018). In addition, an ROI analysis was performed in caudate (CAU), globus pallidus (GP), putamen (PUT), red nucleus (RN), and substantia nigra (SN). A linear regression line and $R^2$ value were calculated between the mean ROI values of QSMnet$^+$ and COSMOS QSM. The results were compared with those of QSMnet.

**Results**

The histogram of the magnetic susceptibility in the training data of QSMnet is plotted in Figure 2. The brain susceptibility distribution of the healthy volunteers ranges from -0.41 ppm to 0.61 ppm whereas 98% of the susceptibility values exist from -0.06 ppm to 0.10 ppm. Top 1% of the susceptibility values range from 0.10 ppm to 0.61 ppm whereas bottom 1% ranges from -0.41 ppm to -0.06 ppm (Fig. 2b), reporting a wider range and higher values of positive susceptibility than negative susceptibility. Compared to this distribution, the susceptibility value of hemorrhagic lesions was much higher (approximately 1.40 ppm; Chang et al., 2016). Therefore, hemorrhagic lesions can be considered as out-of-distribution data in QSMnet.

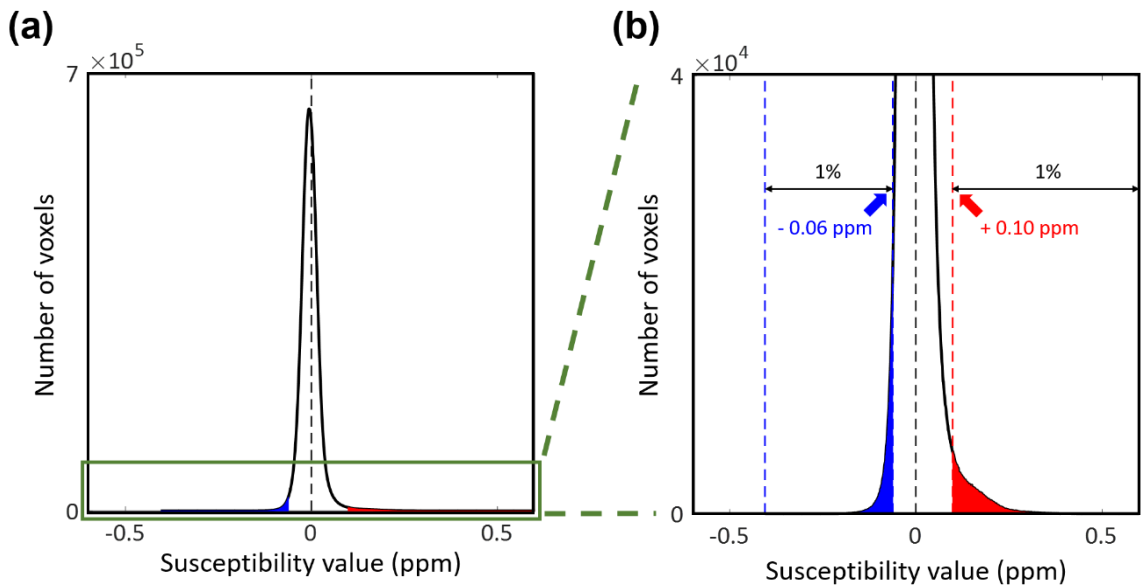

*Figure 2. Training data distribution of the QSMnet susceptibility maps (left) and zoomed-in distribution (right) are plotted. The blue area represents the bottom 1% of the susceptibility distribution and the red area indicates the top 1% of the susceptibility distribution. The black dashed line is a vertical line for zero susceptibility. The susceptibility distribution shows asymmetry with a wider range and higher values of positive susceptibility than negative susceptibility.*

The effect of this out-of-distribution data on QSMnet is shown in Figure 3. When the image from a patient with large hemorrhage is reconstructed using STAR-QSM and QSMnet, the QSMnet result shows superior image quality with little streaking artifacts when compared to that of STAR-QSM (Fig. 3a). However, the susceptibility estimation in the lesion of the QSMnet map is underestimated (Fig. 3b). This result suggests that QSMnet may fail to scale susceptibility values linearly when they are outside of the training data distribution (i.e. -0.41 to 0.61 ppm).

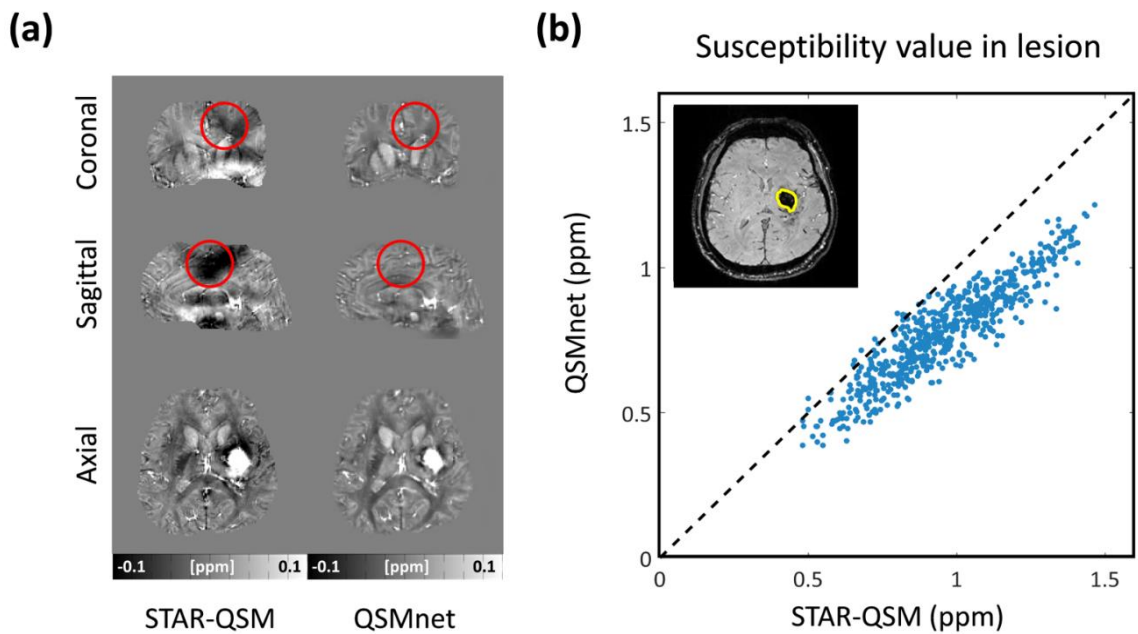

*Figure 3.* (a) Susceptibility maps from STAR-QSM (left column) and QSMnet (right column) in a hemorrhagic patient. The red circles in the STAR-QSM maps indicate streaking artifacts which are not noticeable in the QSMnet maps. (b) Scatter plot of the susceptibility values in the lesion (yellow circle in the inset) generated by STAR-QSM as x-axis and QSMnet as y-axis. The susceptibility values in the QSMnet map are underestimated when compared to those of the STAR-QSM map.

This issue of linearity and improvement using our data augmentation method are demonstrated in the simulated lesion study shown in Figure 4. The plot in Fig. 4a shows the mean susceptibility values of the simulated lesion reconstructed by QSMnet (blue), $QSMnet^+_{\lambda=1}$ (green), $QSMnet^+_{\lambda=2}$ (yellow), $QSMnet^+_{\lambda=3}$ (purple), and $QSMnet^+_{\lambda=4}$ (red). The QSMnet results yield underestimated susceptibility estimation in both positive and negative susceptibility values with larger underestimation in negative susceptibility. This asymmetry may originate from the asymmetric training data distribution (Fig. 2). On the other hand, the QSMnet[+] results, which applied the proposed data augmentation, show less asymmetry. The estimated susceptibility values become closer to the assigned values as λ increases from 1 to 4, demonstrating a successful improvement of linearity (RMSE in QSMnet: 0.36 ppm, $QSMnet^+_{\lambda=1}$: 0.19 ppm, $QSMnet^+_{\lambda=2}$: 0.12 ppm, $QSMnet^+_{\lambda=3}$: 0.07 ppm, and $QSMnet^+_{\lambda=4}$: 0.04 ppm). In Figure 4b, exemplary susceptibility maps from STAR-QSM, QSMnet, and QSMnet[+] are shown for two different susceptibility values assigned to the lesion (left: -1.4 ppm; right: +1.4 ppm). Compared to the results of STAR-QSM and QSMnet, the QSMnet[+] maps show less noticeable artifacts, revealing well-defined lesion boundary. Hence, QSMnet[+] improves not only the linearity of susceptibility values but also image quality.

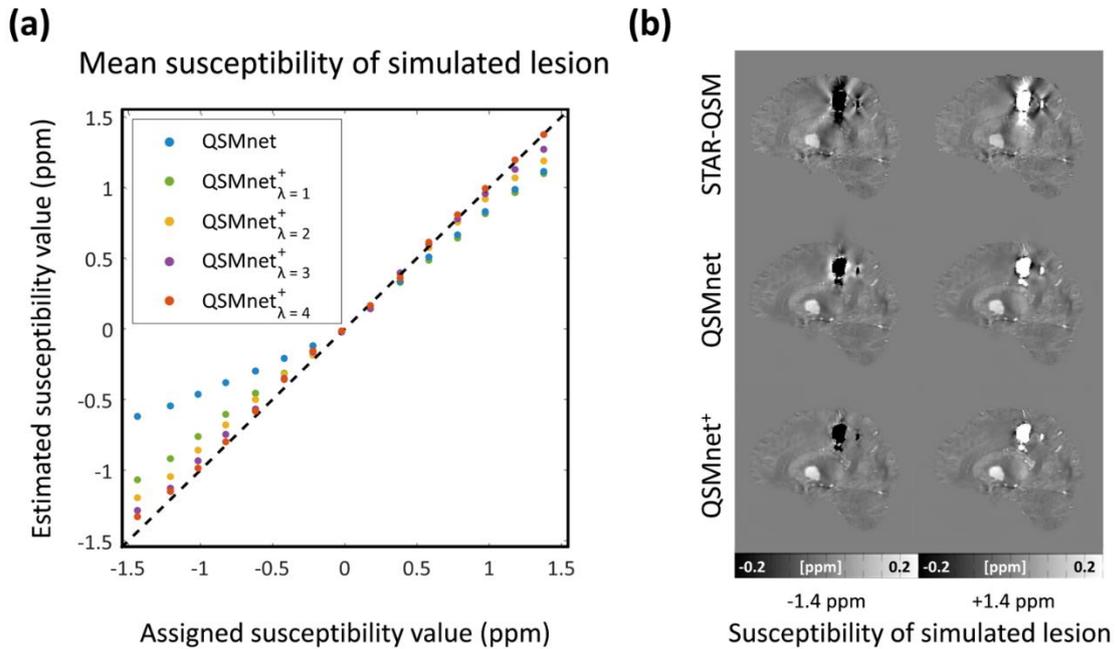

*Figure 4. (a) Assigned value vs. ROI-averaged susceptibility values of lesions reconstructed from each network (QSMnet: blue, QSMnet[+] with scaling factor 1: green, 2: yellow, 3: purple, and 4: red). The mean susceptibility values of the lesion are varied from -1.4 ppm to 1.4 ppm in the step size of 0.2 ppm. The results of $QSMnet^+_{\lambda=4}$ is close to the line of unity (a dashed black line). (b) Reconstruction maps of STAR-QSM, QSMnet, and QSMnet[+] are compared for the simulated lesions with the susceptibility of -1.4 ppm (left) and +1.4 ppm (right).*

The same trends as in the simulated results are observed when the networks are applied to the hemorrhage patients (Figure 5). Magnitude images and four QSM maps reconstructed by MEDI, STAR-QSM, QSMnet, and QSMnet[+] are displayed for four representative hemorrhage patients. Significant streaking artifacts are observed in the MEDI and STAR-QSM maps whereas much reduced artifacts are visible the QSMnet and QSMnet+ maps.

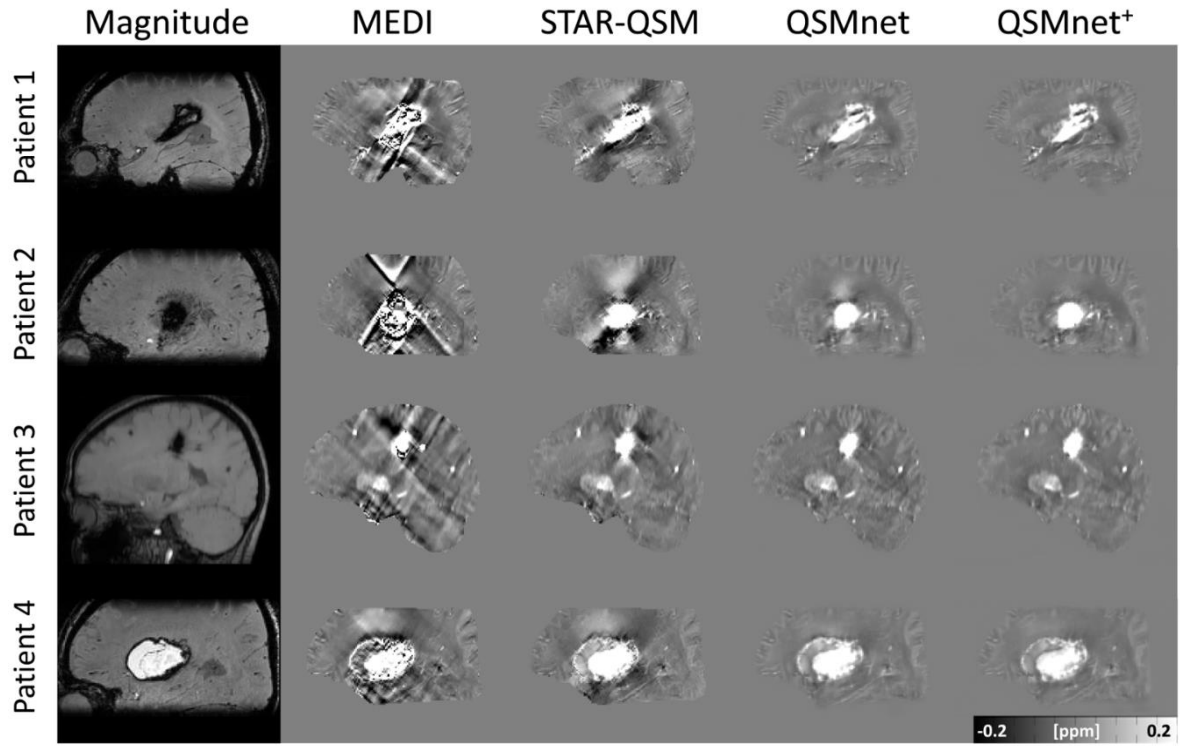

*Figure 5. Magnitude and QSM maps reconstructed by MEDI, STAR-QSM, QSMnet, and QSMnet+ from 4 representative hemorrhagic patients. The streaking artifacts substantially degraded the QSM maps from MEDI and STAR-QSM. On the other hand, the artifacts are less noticeable in the QSMnet and QSMnet+ maps.*

When quantitative analysis is performed for the lesions of the patients, we can reconfirm the improved linearity in the QSMnet+ results over those of QSMnet (Fig. 6). Each dot in Fig. 6a represents the mean susceptibility value of the lesion in a hemorrhagic patient, reconstructed by STAR-QSM and QSMnet (blue dots) or by STAR-QSM and QSMnet+ (red dots). The linearly fitted line of the blue dots (slope = 1.05; intercept = -0.03; $R^2$ = 0.93) is closer to the line of unity than that of the red dots (slope = 0.68; intercept = 0.06; $R^2$ = 0.86). When we focus on one lesion (black circles in Fig. 6a) and generate a scatter plot of the susceptibility values in the lesion, it illustrates that underestimated lesion susceptibility values in QSMnet are successfully corrected in QSMnet+ (Figure 6b). These results consolidate the improved linearity of QSMnet+ in the areas with large susceptibility.

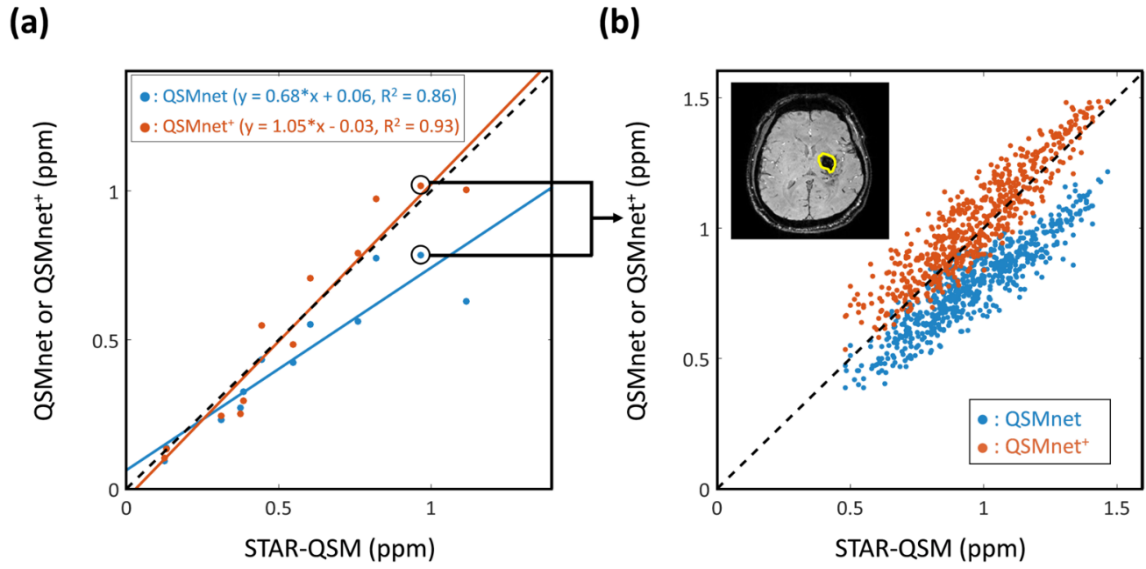

*Figure 6.* Evaluations of QSMnet$^+$ using the hemorrhagic patient datasets. (a) Lesion ROI-averaged susceptibility values from STAR-QSM vs. QSMnet (blue dot) and STAR-QSM vs. QSMnet$^+$ (red dot). Each dot corresponds to each patient. The linear regression line of QSMnet$^+$ (red line) is closer to the line of unity (dashed black line) than that of QSMnet (blue line), suggesting a good correspondence between the two measurements. (b) Scatter plot of the susceptibility values in the lesion in a representative patient noted by the black circles in (a). The susceptibility values (STAR-QSM vs. QSMnet: blue dot, STAR-QSM vs. QSMnet$^+$: red dot) in the lesion ROI (yellow circle in the inset) suggest that the linearity is successfully restored in QSMnet$^+$ in the hemorrhagic patient.

When the five healthy subject data are reconstructed by QSMnet and QSMnet$^+$, all the ROI measurements show a good agreement with those of COSMOS (see Supplementary Figure S5 and Table S2). These results suggest that the proposed data augmentation method does not degrade the image quality of the healthy subjects.

**Discussion and Conclusion**

In this study, we demonstrate that the linearity of QSMnet is undermined in the high susceptibility region (e.g., hemorrhagic lesion). To improve the linearity, a data augmentation method that scales the susceptibility values of the training data, is proposed. The newly trained network, QSMnet$^+$, shows improved linearity when tested with a computer-simulated lesion and hemorrhagic patient data, validating the utility of the proposed data augmentation method.

Our data augmentation approach was designed to generate symmetric linearity for both positive and negative susceptibility as demonstrated in Fig. 2 (e.g. QSMnet$^+_{\lambda=1}$; green dots).

This result is in contrast to the outcome of QSMnet in Fig. 2 (blue dots), which shows further degradation of linearity for negative susceptibility. The asymmetry of the plot for positive and negative susceptibility may originate from the training data distribution of QSMnet, which had higher values of positive susceptibility than negative susceptibility, resulting in better linearity for the positive susceptibility. Additionally, $QSMnet^+_{\lambda=4}$ in Fig. 2 (red dots) shows better agreement with the assigned values when compared to $QSMnet^+_{\lambda=1}$ revealing the improvement of the linearity by expanding training data distribution.

The maximum scaling factor in the data augmentation ($\lambda = 4$) was chosen to cover the susceptibility of hemorrhage, which was reported to be 1.4 ppm (Chang et al., 2016). This value agrees with our hemorrhagic patient results, which show the lesion susceptibility range from -0.63 ppm to 1.42 ppm (see Supplementary Figure S6). The wide susceptibility coverage of $QSMnet^+$ also covers the maximum susceptibility in the deep gray matter of Alzheimer's disease (0.14 ± 0.04 ppm; Moon et al., 2016), Parkinson's disease (0.19 ± 0.05 ppm; Barbosa et al., 2015), and Wilson's disease due to copper accumulation (0.18 ± 0.06 ppm; Fritzsch et al., 2014). Additionally, the negative susceptibility of calcification, which was reported to be -1.4 ppm, is also covered in $QSMnet^+$ (Deistung et al., 2013; Straub et al., 2017). Therefore, $QSMnet^+$ may be suitable for most of pathological conditions. However, the susceptibility values of non-biological materials (e.g. aneurysm clip using titanium) may fall outside of the coverage of $QSMnet^+$.

When developing the network, the optimization of the augmented datasets was performed for the scaling map, scaling factor, and dataset composition. Here, we summarized the results of the optimization. More detailed information can be found in Supplementary Information. In our study, the scaling map, $\Lambda$, was scaled only for positive susceptibility voxels (Eq. 3). When the scaling was applied for both positive and negative voxels, the performance of the network was degraded in the simulated lesion (see Supplementary Figure S1). For the optimization of the scaling factor ($\lambda$), larger values (e.g. $\lambda = 10$) were tested during the development. However, the network trained with the large value of the scaling factor failed to reconstruct the QSM maps of healthy volunteers (Supplementary Figure S4 and Table S1). We also tested the effect of training dataset composition. Three different data sizes for the scaled dataset (20, 50, and 80 maps; equal number of scaled data vs. scaled and inverted data) were compared while fixing the number of unscaled dataset (50 maps). When evaluated by the simulated lesion and the test sets of healthy volunteers, the network trained with 50 scaled maps showed the best results (Supplementary Figure S6). The network trained with 20 scaled maps show larger errors in the simulated lesion, while the network trained with 80 scaled maps show larger errors in the test sets of healthy volunteers. The results suggest that the balance between the experimental dataset and scaled dataset is important.

When compared to the original QSMnet in Yoon et al. (Yoon et al., 2018), we improved a few components of the neural network. First, both input and output of the training dataset were normalized to have a mean of 0 and a standard deviation of 1. This normalization has shown to improve training efficiency (LeCun et al., 1998). Secondly, the model loss function was reformulated from L1 difference between the dipole convolution of the label and the dipole convolution of the output (i.e. $\text{loss}_{\text{Model}} = ||(d * y - d * \chi)||_1$) to L1 difference between the dipole convolution of the label and the input local field (Eq. 7; $\text{loss}_{\text{Model}} = ||M \cdot (\delta - d * \chi)||_1$). Although this modification had minor improvements (data not shown), the new model loss conveys better physical meaning of the model. Lastly, the activation function was changed from ReLU to leaky ReLU. When tested using the simulation, the leaky ReLU improved the linearity of QSMnet$^+$ (Supplementary Fig. S9).

Recently, a few different approaches of deep neural network-powered QSM reconstruction have been proposed (Bollmann et al., 2019; Chen et al., 2019a; Chen et al., 2019b; Liu and Koch, 2019a, b; Liu et al., 2019a; Liu et al., 2019b; Polak et al., 2019; Wei et al., 2019; Zhang et al., 2019). Among them, the methods by Polak et al., combined the physical model of dipole and variational networks (Hammernik et al., 2018; Polak et al., 2019), and the work by Zhang et al., developed the loss function based on the physical model of dipole (Zhang et al., 2019). These methods may potentially have improved linearity when compared to the end-to-end trained QSMnet. However, the test of linearity was not performed in the works.

In conclusion, we explore the linearity of a deep neural network in MR parametric mapping and suggested a data augmentation method to improve the linearity. Since a few deep neural network trained MR reconstruction methods target for quantitative parametric mapping (Bertleff et al., 2017; Cohen et al., 2018; Lee et al., 2018), this linearity issue may be an important consideration. Our augmentation method may be applicable for them. Finally, the linearity improved susceptibility mapping, QSMnet$^+$, covers a wider range of susceptibility values, and, therefore, can be used as a tool for clinical studies.

**Acknowledgment**

This work was supported by the National Research Foundation of Korea (NRF) grant funded by the Korea government (MSIT) (NRF-2018R1A2B3008445).